\documentstyle[12pt,epsf]{article}

\newcommand{\e}{\epsilon}

\newcommand{\Ft}{\tilde{F}}
\newcommand{\m}{\mu} 
\newcommand{\n}{\nu} 
\newcommand{\la}{\lambda}
\newcommand{\k}{\kappa}

\newcommand{\p}{\partial}  
 
\newcommand{\be}{\begin{equation}}
\newcommand{\ee}{\end{equation}}

\begin{document}
\title{\bf Non-minimal Maxwell-Chern-Simons theory and the composite Fermion model}
\author{Ricardo C. Paschoal$^{1,2}$ \thanks{E-mail: paschoal@cbpf.br}$\;$ and Jos\'{e} A. Helay\"{e}l-Neto$^{1,3}$ \thanks{E-mail: helayel@cbpf.br}\\ \\
$^{1}${\it  \normalsize Centro Brasileiro de Pesquisas F\'{\i}sicas -- CBPF,} \\
{\it \normalsize Rua Dr. Xavier Sigaud 150, 22290-180, Rio de Janeiro, RJ, Brasil}  \\ 
$^{2}${\it  \normalsize Servi\c{c}o Nacional de Aprendizagem Industrial, } \\
{\it  \normalsize Centro de Tecnologia da Ind\'{u}stria Qu\'{\i}mica e T\^{e}xtil -- SENAI/CETIQT, } \\
{\it \normalsize Rua Dr. Manoel Cotrim 195, 20961-040, Rio de Janeiro, RJ, Brasil} \\  
$^{3}${\it \normalsize Grupo de F\'{\i}sica Te\'{o}rica Jos\'e Leite Lopes,} \\
{\it \normalsize Petr\'opolis, RJ, Brasil}  }

\date{ }

\maketitle 

\begin{abstract} 

The magnetic field redefinition in Jain's composite fermion model for the fractional quantum Hall effect is shown to be effectively described by a mean-field approximation of a model containing a Maxwell-Chern-Simons gauge field non-minimally coupled to matter.
Also an explicit non-relativistic limit of the non-minimal (2+1)D Dirac equation is derived.

\end{abstract} 

\newpage

Since the introduction of the Chern-Simons (CS) term in the Lagrangean of planar Electrodynamics~\cite{DJT82-etc}, it has been used in many aplications, and probably the most successful of them was the establishment of a connection between CS theory and fractional statistics~\cite{anyons}, mainly when applied to the physics of the fractional quantum Hall effect (FQHE)~\cite{FQHE}. Moreover, the two space dimensions open the possibility for a natural inclusion of a non-minimal (Pauli-type) coupling~\cite{Stern91,Kogan91,Lat-Soro-91}\footnote{Even for the case of spin zero (scalar) matter~\cite{Stern91,Kogan91,Lat-Soro-91}.}, as it will be briefly reviewed below. These two terms --- CS and non-minimal coupling --- are induced by radiative corrections~\cite{DJT82-etc,Kogan91} in planar QED, which is a sufficient motivation for including them in the bare Lagrangean.
In addition, however, some interesting connections between these two terms were also observed in the scalar theory:
when a non-minimal coupling is assumed in the Lagrangean, a CS term is generated by spontaneous symmetry  breaking~\cite{PaulKhare87}
and fractional statistics may be achieved even without an explicit CS term, either in the non-relativistic limit~\cite{Carr-Kuns-95} or in the relativistic one~\cite{Nobre-Almeida-99}; 
a CS counter-term cures (at one-loop\footnote{At two loops, the renormalizability is still kept, but only for a certain value of the non-minimal coupling constant~\cite{Carr-Kuns-99}.}) the non-renormalizability of a non-minimal term~\cite{Carr-Kuns-94}; and, as well,  a non-minimal coupling cures~\cite{KimLeeYee-99}, at least in the non-relativistic limit, the singularity problem of the scalar Maxwell-Chern-Simons (MCS) theory, which were pointed out before, by Ref.~\cite{Nemeth98}.
Not restricted to the scalar theory, but also valid in the fermion one, other important results are the 
vortex solutions in MCS gauge theory, both in Abelian and non-Abelian cases, and both relativistically or not, which were made possible by the inclusion of a non-minimal coupling~\cite{KimLeeYee-99}.

Another motivation for considering a non-minimal MCS theory is the success of Jain's composite fermion (CF) model for the FQHE~\cite{Jain}, in which each electron is assumed to have an even number of magnetic flux quanta (or {\it fluxons}) attached to it, an idea brought up by the connection between CS theory and statistics transmutation mentioned above. A CF is a ``particle" with charge and flux. Since, in two space dimensions, the magnetic flux and the magnetic dipole moment have the same dimension ($=$ mass$^{-1/2}$ in a $\hbar=c=1$ unit system)\footnote{It is worth stressing that this is a peculiarity of two space dimensions. It is not valid in (1 + 3) dimensions, where the magnetic flux is mass$^0$ and the magnetic dipole moment is mass$^{-1}$.  Indeed, the main result of the present work --- built up in two space dimensions --- is in accordance with the idea of identifying these two quantities.},
a CF may be viewed as a ``particle" with both charge and magnetic dipole moment, just as in a Pauli-type non-minimal coupling scheme and thus some similarity between these two models would not be much surprising. Indeed, a relation between the CS term and a fluid of particles having a charge and a magnetic moment was stablished in Ref.~\cite{BarciOx95}. That work also showed the interaction between the currents of such particles and the gauge field to be equivalent --- up to a surface term --- to a Pauli-type one.

In order to implement a non-minimal coupling, one first notes that while in (1+3) dimensions the dual $\Ft^{\m\n} \equiv \frac{1}{2}\e^{\m\n\k\la}F_{\k\la}$ of the electromagnetic field $F_{\k\la}$ is a second-rank tensor, on the other hand, in (1+2) dimensions, it is a vector, $\Ft^\m \equiv \frac{1}{2}\e^{\m\k\la}F_{\k\la}$, and thus the minimal covariant derivative 
\be
        D_\m  \equiv  \p_\m + iqA_{\m} 
\ee 
may be generalized to a non-minimal one,
\be
        D_\m  \equiv  \p_\m + iqA_\m + ig\Ft_{\m}  , \label{cov-deriv}
\ee
where $g$ is the planar analogue of the magnetic dipole moment, which couples non-minimally with the magnetic field\footnote{Notice that the derivative (\ref{cov-deriv}) behaves like a covariant one, since the non-minimal term is gauge covariant, by definition.}. Indeed, since $\Ft^\m = (-B, -E_y, E_x) \equiv (-B, -\tilde{\vec{E}})$, the equation above splits in components as:
\be
        D_0  \equiv  \frac{\p}{\p t} + iq\Phi - igB 
\ee
and
\be
       D_i \equiv  \p_i - iq(\vec A)_i + ig\tilde{E}_{i} .
\ee

To obtain the Schr\"odinger equation for an electron subject to an electromagnetic field, one proceeds as usual, substituting $\p_0=\p_t$ with $D_0$ (or, equivalently, summing the term $q\Phi -gB$ to the Hamiltonian) and $\p_i$ with $D_i$, these substitutions being readily seen to be equivalent to the minimal prescription, except for the following changes: 
\be
       \Phi \rightarrow \Phi -\frac{g}{q}B
\ee
and
\be
       (\vec{A})_i \rightarrow (\vec{A})_i - \frac{g}{q}\tilde{E}_i ,
\ee
which, due to the definitions\footnote{It is worthwhile to remark that Eq.~(\ref{def-B}) is a scalar one.}:
\be
     B\equiv\vec\nabla\times\vec{A}\equiv\epsilon_{ij}\p_i(\vec{A})_j      \label{def-B}
\ee
and
\be
     \vec E \equiv  \frac{\p}{\p t}(-\vec{A}_i) - \vec{\nabla}_i \phi ,
\ee
imply that:
\be
       B \rightarrow B' = B + \frac{g}{q}\left( \vec{\nabla}\cdot\vec{E} \right) \label{redefB}
\ee
and
\be
       \vec E \rightarrow \vec E' = \vec E + \frac{g}{q}\left( \vec{\nabla}B + \frac{\p\tilde{\vec E}}{\p t}  \right), \label{redefE}
\ee
the Lorentz force preserving its form.

Performing the substitutions described above, one easily obtains the Schr\"odinger equation for the electron wave function $\psi(x, y, t)$:
\be
         i\frac{\p \psi}{\p t} = \left\{ q\Phi + \frac{ \left( \vec{p} - q\vec{A} + g\tilde{\vec{E}} \right)^2 }{2M} - gB \right\} \psi, \label{Sch-n-min}
\ee
where $M$ is the electron mass, $\vec p = -i\vec\nabla$ is the canonical momentum and  $-gB$ plays the role of a magnetic dipole energy. The term involving $g\tilde{\vec{E}}$ was shown in Ref.~\cite{Carr-Kuns-95} to give rise to an Aharonov-Casher phase in the scalar theory\footnote{Besides the Aharonov-Bohm phase generated  by the $-q\vec A$ term, as usual.}.
Now, a good issue is to write down the (2+1)-dimensional Dirac equation using the same non-minimal covariant derivative and verify whether its non-relativistic limit differs from equation above by some extra term(s), as it occurs with Pauli term in the minimal (3+1)-dimensional well known case~\cite{Bjorken-Drell-64}. This question is already positively answered in Refs.~\cite{Stern91,Kogan91,Lat-Soro-91}, but rather exploring the peculiar properties of the (2+1)-dimensionality than by means of the explicit use of the (2+1)-dimensional Dirac equation, which seeems to be lacking in the literature and will be performed below\footnote{In Ref.~\cite{Carr-Kuns-95}, this procedure is undertaken  for the scalar theory but the fermion field is considered only after a dimensional reduction of the (3+1)-dimensional Dirac equation.}.

From Dirac equation, $\left( i\gamma^\m D_\m - M \right)\Psi = 0$, with the covariant derivative (\ref{cov-deriv}) and a two-component Dirac spinor  $\Psi = ( \varphi'\:\chi' )^T$ (in view of the three spacetime dimensions), one obtains:
\be
      i\frac{\p}{\p t}\left( \begin{array}{c}
                                    \varphi \\ \chi
                                    \end{array} \right) =
      \left( q\Phi - gB \right) \left( \begin{array}{c}
                                    \varphi \\ \chi
                                    \end{array} \right)  -
     2M\left( \begin{array}{c}
                                    0 \\ \chi
                                    \end{array} \right) +
      \tilde\pi_x\left( \begin{array}{c}
                                    \chi \\ \varphi
                                    \end{array} \right)  +
      i\tilde\pi_y\left( \begin{array}{c}
                                   - \chi \\ \varphi
                                    \end{array} \right)           , \label{BD-1.30}
\ee
where $(\varphi'\:\chi')=e^{-imt}(\varphi\:\chi)$,  $\tilde{\vec{\pi}} = ( \pi_y, - \pi_x )$ and $\vec{\pi}=\vec p - q\vec A + g\vec E$ is the kinetic momentum. The following representation of the gamma matrices was used:   $\gamma^0 = \sigma_z$,  $\gamma^1 = i\sigma_x$ and $\gamma^2 = i \sigma_y$, the $\sigma$'s being the usual $2\times 2$ Pauli matrices.

In the non-relativistic and weak field limits ($2M,\:\tilde\pi_i \gg i\p_t,\: q\Phi,\: gB$), the second equation above reads:
\be
         \chi = \frac{\tilde\pi_+ \varphi}{2M}, \label{BD-1.30a}
\ee
where $\tilde\pi_\pm \equiv \tilde\pi_x \pm i\tilde\pi_y$, showing that $\chi$ is negligible compared to $\varphi$. Substituting (\ref{BD-1.30a}) in the first component of Eq.~(\ref{BD-1.30}), one obtains:
\be
            i\frac{\p \varphi}{\p t} =  \left( q\Phi -gB + \frac{\pi_-\pi_+}{2M} \right) \varphi   ,
\ee
which, after the proper substitutions and using $[p_i,\: f] = -i(\p_i f)$, yields the non-relativistic limit of the (2+1)-dimensional Dirac equation with a non-minimal coupling to an Abelian vector gauge field:
\be
      i\frac{\p \varphi}{\p t} =
      \left\{ q\Phi + \frac{ \left( \vec p - q\vec{A} + g\tilde{\vec{E}} \right)^2 }{2M} - \frac{qB}{2M} - gB 
      - \frac{g}{2M}\left( \vec{\nabla}\cdot\vec{E} \right) \right\} \varphi    .      
      \label{2+1Dir-nrel-nmin}       
\ee

In the Hamiltonian inside the brackets of Eq.~(\ref{2+1Dir-nrel-nmin}), one recognizes the first term as the usual electric charge-field energy; the second one is a generalization of the usual kinetic energy, with the kinetic momentum now given by
$\vec p - q\vec{A} + g\tilde{\vec{E}}$; the third and fourth terms respectively are the (2+1)-dimensional Pauli term and a tree-level anomalous magnetic moment contribution as described in Refs.~\cite{Stern91,Kogan91,Lat-Soro-91}; and the last one as an electric {\em quadrupole} moment interaction, which  deserves deeper interpretation, but is indeed much more surprising (for a point-like particle) than the magnetic dipole term itself, once for the latter there exists the appeal to the particle spin as the usual interpretation\footnote{Although even this argument fails in the scalar case, for which the anomalous magnetic moment term remains, as already mentioned above.}. Equation above is equal to the first component of the 2-component spinor Eq.~(21) in Ref.~\cite{Carr-Kuns-95}, derived from the non-relativistic limit of the (3+1)-dimensional (4-component) Dirac equation with a Pauli term, by means of a dimensional reduction.
However, the equation above seems more natural to describe a planar particle, which does not have two spin degrees of freedom, but just one. Nevertheless, it is interesting to know that the directly (2+1)-dimensional approach adopted here can describe essentially the same physics as the dimensional reduction of the ``real" (3+1)-dimensional world.

The gauge field, $A_\mu$, will be assumed to be governed by a MCS action, which generates the following Euler-Lagrange field equations~\cite{DJT82-etc}:
\begin{eqnarray}
 & & \vec{\nabla}\cdot\vec{E}-mB=\rho \label{MCS1} \\ 
 & & \vec{\nabla}\times\vec{E}= - \frac{\p B}{\p t} \label{MCS2} \\
 & & \tilde{\vec{\nabla}}B - m\tilde{\vec{E}} = \vec{J} +  \frac{\p \vec{E}}{\p t}\label{MCS3},
\end{eqnarray}
where:
\begin{eqnarray}
\tilde{\vec{\nabla}}_i & \equiv & \e_{ij}\vec{\nabla}_j = \e_{ij}\p_j \\
\tilde{\vec{E}}_i & \equiv & \e_{ij}\vec{E}_j = - \Ft^i = \Ft_i \\
J^\m & = & (\rho, \vec{J}).
\end{eqnarray}

These equations do not take into account the electromagnetic fields generated by the magnetic dipoles $g$. This is an approximation which is in accordance with the fact that the applied magnetic field in a Hall bar is much stronger than those fields\footnote{Otherwise, the Lagrangian and the corresponding field equations would be respectively~\cite{Georg-Wall-92}:
${\cal L} = -\frac{1}{4}F_{\mu\nu}^2 + \frac{m}{2}\e^{\mu\k\lambda}A_{\m}\p_\k A_\la - J_\m A^\m - \mu\e^{\mu\k\lambda}J_{\m}\p_\k A_\la$;
$\vec{\nabla}\cdot\vec{E}-mB=\rho - \mu \vec{\nabla}\cdot\tilde{\vec{J}}$;
$\vec{\nabla}\times\vec{E}= - \frac{\p B}{\p t}$; and
$\tilde{\vec{\nabla}}B - m\tilde{\vec{E}} = \vec{J} +  \frac{\p \vec{E}}{\p t} + \mu\left( \tilde{\vec{\nabla}}\rho + 
\frac{\p \tilde{\vec{J}}}{\p t} \right)$, where $\mu\equiv g/q$.
}.
Other approximations will be made, all of them consistent with the mean fields inside a Hall bar: it will be assumed $\rho=0$ (locally neutral specimen) and magnetic and electrical fields constant with time. Hence
\begin{eqnarray}
 & & \vec{\nabla}\cdot\vec{E} = mB \label{MCS1a} \\ 
 & & \vec{\nabla}B  =  m\vec{E} - \tilde{\vec{J}}  \label{MCS3a}.
\end{eqnarray}

In a Hall bar, electric and magnetic fields are also uniform,  but the unconventional MCS equation (\ref{MCS1a}) does not allow it to be true with respect to the electric field. Indeed, the connection between charge and magnetic flux (which is more evident after an area integration of (\ref{MCS1}) ) is a well-known characteristics of (M)CS theory, and both the magnetic field and electrical charge may be viewed as a font of $\vec{E}$.  However, Eq.~(\ref{MCS3a}) enables the magnetic field  to be kept as uniform, provided the condition $m\vec{E} = \tilde{\vec{J}}$ is satisfied, which is nothing but the identification of the CS parameter $m$ with the Hall conductivity, another well-known result of CS theory, which is thus also valid in MCS Electrostatics (plus a uniform magnetic field).

The conclusion is that, in a Hall bar, a particle which possesses both an electric charge, $q$, and a magnetic dipole moment, $g$, subject to electric and magnetic fields $\vec{E}$ and $B$, may be thought of as a particle with the same charge, $q$, but no magnetic dipole moment, `feeling'  the following modified fields of Eqs.~(\ref{redefB}) and (\ref{redefE}):
\be
       B' = B + \frac{g}{q}\left( \vec{\nabla}\cdot\vec{E} \right) = B \left( 1 + \frac{gm}{q} \right) \label{redefBa}
\ee
and
\be
       \vec E' =  \vec E,  \label{redefEa}
\ee
where Eqs.~(\ref{MCS1a}) and (\ref{MCS3a}), along with $\vec{\nabla} B = 0$, were used.

Now, a similar result of Jain's CF model for the FQHE~\cite{Jain} will be mentioned: it also refers to a redefinition of the magnetic field: 
\be
       B^* = B_0 - 2n\rho_e \phi_0  \label{redefB-Jain},
\ee
where $B_0$ is the external magnetic field, experienced by the electron; $B^*$ is the effective magnetic field seen by the CF (an electron with an even number of magnetic fluxons attached); $n$ is a positive integer; $\rho_e$ is the mean electron (area) density; and $\phi_0$ is the fluxon value (or the magnetic flux quantum), $\phi_0=hc/e$, or $\phi_0=2\pi/e$ in $\hbar=c=1$ unit system.

Since a CF is endowed with both an electric charge and a magnetic flux (which, as already pointed out, in (1+2) dimensions has the same unit as a magnetic dipole moment), an analogy emerges between the quantities $\{B_0, B^*\}$ and respectively the quantities $\{B', B\}$ in Eq.~(\ref{redefBa}) above. Indeed, from Eqs.~(\ref{redefB-Jain}) and (\ref{redefBa}), the equality between these two pairs of quantities is possible if one assumes that ($q=-e$):
\be
         m = - \frac{4\pi\rho_e}{Bg}n. \label{JainMCS}
\ee

An important quantity in the QHE is the so-called filling factor $\nu_e = \rho_e\phi_0/B_0$ (or, identifying $B_0$ with $B'$ and using $\phi_0 = 2\pi / e$, $\nu_e = 2\pi\rho_e/eB'$). It corresponds to the number of occupied Landau levels, or, more precisely, the ratio between the number of electrons and the degeneracy of each Landau level. An equivalent quantity is defined for the CFs~\cite{Jain}: $\nu_{CF} = \rho_e\phi_0/B^*$ (or $\nu_{CF} = 2\pi\rho_e/eB$), where $\nu_{CF}$ and $B^*$ (or $B$, as proposed here) may be both positive, or both negative, depending respectively whether the magnetic fields $B_0$ and $B^*$ are parallel or antiparallel. From these two definitions and Eq.~(\ref{redefB-Jain}), a relation is derived~\cite{Jain} between $\nu_e$ and $\nu_{CF}$:
\be
        \nu_e = \frac{\left|\nu_{CF}\right|}{2n\left|\nu_{CF}\right| \pm 1}, \label{nue-nuCF}
\ee
where the upper (lower) sign corresponds to $B_0$ and $B^*$ being parallel (antiparallel).

Now consider the system of CFs with quantized Hall conductivity, i.e.,
\be
       m = \frac{e^2}{h}n^* = \frac{e}{\phi_0}n^*, \label{quantiz-m}
\ee
with $n^*$ a positive integer, which is nothing but the absolute value of the CF filling factor: $\left|\nu_{CF}\right|=n^*$. This quantization may be accomplished by means of, for example, Laughlin's argument~\cite{Laugh81} applied to the CF system. In this case, Eq.~(\ref{nue-nuCF}) gives:
\be
        \nu_e = \frac{n^*}{2nn^* \pm 1} \label{nue-n'},
\ee
which coincides with almost all experimentally observed values of the (electron) Hall conductivity $\sigma_H=(e^2/h)\nu_e$ in the fractional QHE, constituting one of the successes of Jain's model. Because of this relation, the CF model is known as a description of the fractional QHE of the (interacting) electrons as an {\it integer} QHE of the (free) CFs. The possible equivalence between this feature of Jain's model and the non-minimal MCS model described above (in mean field approximation), expressed in proposition (\ref{JainMCS}), is enforced by the fact that, in the case of the fractional QHE (i.e., integer QHE for the CFs, $m = (e/\phi_0)n^*$), Eq.~(\ref{JainMCS}) leads to:
\be
         g = \mp 2n\phi_0,      \label{main-result}
\ee
in agreement with the idea in Jain's model that an even number of fluxons corresponds to the value of the CF magnetic dipole moment. Of course, the path undertaken until here could be inverted and Jain's redefinition (\ref{redefB-Jain}) would be derived from Eq.~(\ref{main-result}). In other words: one assumes $g = \mp 2n\phi_0$  in the non-minimal covariant derivative (\ref{cov-deriv}), as a description of particles with charge and magnetic moment (the CFs). Then, assuming that the gauge field, $A^\m$, is governed by MCS equations in a mean field approximation (Eqs.~(\ref{MCS1a}-\ref{MCS3a})), the redefinition of the magnetic field given by Eq.~(\ref{redefBa}) leads to $B'=B \mp 2n\phi_0 m B / q$, where $B'$ is the magnetic field as seen by particles with the same charge as the CFs, but with no fluxons attached (thus, the electrons themselves). Finally, one uses Eq.~(\ref{MCS3a}) in the static case and with a uniform magnetic field to identify the CS parameter $m$ with the Hall conductivity, which, if quantized as in Eq.~(\ref{quantiz-m}), leads, along with $q=-e$ and
$n^*=\left| \nu_{CF} \right| = \rho_e \phi_0 / \left| B \right|$, to Jain's equation (\ref{redefB-Jain}), with $B$ ($B'$) playing the role of $B^*$ ($B_0$).

It is noticeable that the interesting result above has been obtained with very simple (mean field) approximations on the fields inside a Hall bar. However, it would not be attainable if a single of the following assumptions made here were not satisfied: (i) planar physics (which permits the identification of the dimensions of $g$ with that of $\phi_0$, and allows for the existence of the CS term); (ii) CS term (if $m=0$, Eq.~(\ref{redefBa}) does not redefine the magnetic moment as in Jain's model); (iii) non-minimal coupling  (idem for $g$); (iv) Maxwell term (in order to the divergence of the electrical field to be present in Eq.~(\ref{MCS1})).
Of course, additional investigations are needed in order to discover to which extent the non-minimal MCS Electrodynamics yields correct results for the FQHE. One of them, already under study, is to relax the restrictions concerning the fields and consider the flutuactions in $\rho$, by means of a full quantum field theory, assuming (for example) Dirac (spinor) matter fields. Another calculation, not related to the QHE, but which may yield useful results is the electromagnetic field produced by an accelerated charge, in a non-minimal coupling, classical MCS theory. Some interesting results, such as violation of Huygens principle, were already obtained in~\cite{Winder} under a minimal coupling scheme. It will be studied whether the non-minimal coupling introduces new features. Probably a kind of dipole field will arise. Finally, we mention that more quantitative data, concerning the {\it minimal} MCS classical theory, will be published in Ref.~\cite{Eu-H-Fern}.

\section*{Acknowledgements}

We thank A.L.M.A. Nogueira and D. Barci for fruitful discussions and H.~Belich Jr. and M.~Messias Jr. for reading the manuscript.

\end{document}